\documentclass[11pt]{article}

\usepackage[utf8]{inputenc}
\usepackage[T1]{fontenc}
\usepackage{mathptmx}
\usepackage[scaled=0.92]{helvet}
\usepackage{courier}
\usepackage{microtype}

\usepackage[
  top=1in,
  bottom=1in,
  left=1.5in,
  right=1.5in
]{geometry}

\usepackage{fancyhdr}
\fancypagestyle{plain}{\fancyhf{}\fancyfoot[C]{\thepage}}

\usepackage[colorlinks=true,linkcolor=blue!70!black,citecolor=blue!70!black,urlcolor=blue!70!black]{hyperref}
\usepackage{url}

\usepackage{booktabs}

\usepackage{listings}
\usepackage{xcolor}
\usepackage{parskip}
\usepackage{titlesec}
\titleformat{\section}{\large\bfseries}{\thesection}{1em}{}
\titleformat{\subsection}{\normalsize\bfseries}{\thesubsection}{1em}{}
\titlespacing*{\section}{0pt}{1.5em}{0.5em}
\titlespacing*{\subsection}{0pt}{1em}{0.3em}

\usepackage{authblk}

\title{\Large\bfseries Lore: Repurposing Git Commit Messages as a\\Structured Knowledge Protocol for AI Coding Agents}

\author{Ivan Stetsenko}
\affil{Independent Researcher \\ {\small \href{mailto:dev.ivan.stetsenko@gmail.com}{dev.ivan.stetsenko@gmail.com}}}

\date{March 2026}

\begin{document}

\maketitle
\thispagestyle{plain}

\begin{abstract}
\noindent As AI coding agents become both primary producers and consumers of source code, the software industry faces an accelerating loss of institutional knowledge. Each commit captures a code diff but discards the reasoning behind it---the constraints, rejected alternatives, and forward-looking context that shaped the decision. I term this discarded reasoning the \emph{Decision Shadow}. Prior work has identified this tacit knowledge gap as a critical bottleneck for AI-assisted development~\cite{peng2025cdt, terragni2025future} and has proposed heavyweight infrastructure to address it, including knowledge graphs~\cite{peng2025cdt}, version-controlled agent memory systems~\cite{gcc2025}, and separate Architecture Decision Records~\cite{nygard2011}. This paper argues that the solution already exists in every software project: the git commit message. I propose \textbf{Lore}, a lightweight protocol that restructures commit messages---using native git trailers---into self-contained decision records carrying constraints, rejected alternatives, agent directives, and verification metadata. Lore requires no infrastructure beyond git, is queryable via a standalone CLI tool, and is discoverable by any agent capable of running shell commands. This paper formalizes the protocol, compares it against five competing approaches, stress-tests it against its strongest objections, and outlines an empirical validation path. To the best of my knowledge, Lore is the first proposal to repurpose the commit message itself as a structured, machine-parseable knowledge channel optimized for inter-agent communication across time.
\end{abstract}

\section{Introduction}

\subsection{The Decision Shadow}

Every commit in a software project is the visible output of an invisible process. A developer (or AI agent) encounters a problem, considers several possible approaches, evaluates the tradeoffs, selects one, and implements it. The commit captures exactly one artifact from this process: the final diff. Everything else---the problem definition, the alternatives considered, the reasons for rejection, the constraints that shaped the decision, the confidence level, the known weaknesses---evaporates.

I call this lost context the \textbf{Decision Shadow}: the unrecorded reasoning behind why the code looks the way it does at any given point.

Over time, Decision Shadows accumulate. Each one is individually small. Collectively, they produce what the industry calls ``legacy code''---code that functions but whose structural rationale is lost. Peng and Wang~\cite{peng2025cdt} describe this phenomenon as ``tacit knowledge'' that ``often lives in developer experience or informal artifacts rather than in code'' and observe that ``current [AI] assistants cannot reliably retrieve or reconstruct this knowledge on demand.''

\subsection{The Commit Message as It Exists Today}

The Conventional Commits specification~\cite{conventionalcommits}, the closest approximation to an industry standard, encodes a message as \texttt{type(scope): short description}. This tells you what happened (\texttt{fix(auth): handle expired token refresh}) but is nearly useless for understanding \emph{why} the code evolved the way it did.

The problem is structural, not motivational. The format was designed for a world where humans were the only consumers of commit history, the diff was the primary artifact, and deep context lived in people's heads---transmitted orally through standups, pull request reviews, and hallway conversations. All three assumptions are collapsing.

\subsection{Why This Matters Now}

Two shifts make the Decision Shadow problem urgent.

\textbf{Shift 1: AI agents are now primary code consumers.} Tools such as Claude Code~\cite{claudecode}, GitHub Copilot~\cite{copilot}, and Cursor~\cite{cursor} attempt to reconstruct intent and constraints from existing code when tasked with modifications. The code tells them what exists; tests tell them what should be true; the commit history should tell them \emph{why}---but in practice, messages like \texttt{refactor: clean up utils} transmit near-zero signal.

\textbf{Shift 2: AI agents are now primary code producers.} When an agent generates a commit, the message it writes is typically a summary of the diff---a lossy compression of information already present. Research from GitClear analyzing 153 million changed lines of code found that AI coding tools are introducing code faster than ever while raising concerns about long-term maintainability~\cite{gitclear2024}. No decision context is added to commits. The knowledge destruction problem is \emph{accelerating}.

The compound result: codebases are growing faster than ever while the ratio of recorded institutional knowledge to code volume is decreasing.

\section{Related Work}

\subsection{Architecture Decision Records}

Architecture Decision Records (ADRs), popularized by Nygard~\cite{nygard2011} in 2011, capture high-level architectural decisions as standalone documents. The format typically includes a title, context, decision, and consequences. ADRs have gained adoption in government~\cite{18f2021} and enterprise settings~\cite{haylock2025}, and tools such as Log4brains~\cite{log4brains} provide management and publication infrastructure.

ADRs operate at the architectural granularity---explaining choices such as ``why we chose PostgreSQL over MongoDB''---but do not capture implementation-level decisions: why a particular function handles errors the way it does, what alternatives were considered for a specific module's retry logic, or what constraints shaped a data validation approach. These implementation decisions are far more numerous than architectural decisions and are the ones most commonly lost. Furthermore, ADRs suffer from a synchronization problem: as separate files, they must be manually maintained alongside an evolving codebase and routinely go stale~\cite{18f2021}. Lore addresses the complementary granularity---implementation decisions---and eliminates the synchronization problem by embedding knowledge directly in the commit, which is immutable and atomically bound to the code change.

\subsection{Code Digital Twin}

Peng and Wang~\cite{peng2025cdt} propose the Code Digital Twin, a ``living knowledge infrastructure that couples a physical layer of software artifacts with a conceptual layer that encodes system intent and evolution.'' Their diagnosis of the problem aligns strongly with the one presented here: tacit knowledge including ``architectural rationales, design trade-offs, and historical context'' is critical for AI-assisted development, and ``capability gains alone do not eliminate the need for structured, version-aware context that preserves why the system is the way it is.''

However, their proposed solution is architecturally heavyweight: a multi-layered infrastructure involving knowledge graphs, bidirectional traceable links between code artifacts and conceptual models, and continuous automated extraction pipelines. While this may be appropriate for large enterprise systems such as the Linux kernel (their motivating example), it represents significant infrastructure investment that most projects cannot undertake. Lore identifies the same problem but proposes a radically lighter-weight treatment: enriching an artifact every project already has (commit messages) using a mechanism git already supports (trailers), queryable through a tool every agent already knows how to use (a CLI).

\subsection{Git Context Controller}

The Git Context Controller (GCC)~\cite{gcc2025} proposes a structured context management framework for AI agents, using git-inspired operations (COMMIT, BRANCH, MERGE, CONTEXT) over a \texttt{.GCC/} directory to organize agent memory as a versioned file system. The authors report state-of-the-art results on the SWE-Bench benchmark when agents are equipped with GCC.

GCC and Lore address different aspects of the agent context problem. GCC is an \emph{intra-session} memory management system: it helps an agent organize its own working memory during a task, with checkpointing and branching to support exploration. Lore is an \emph{inter-session} knowledge transfer mechanism: it encodes decision context into the permanent project history so that \emph{future} agents (or humans) benefit from it. GCC manages the agent's scratchpad; Lore manages the project's institutional memory. The two are complementary---an agent could use GCC during a task and produce a Lore-enriched commit at the end.

\subsection{AI Code Attribution and Provenance}

The git-ai project~\cite{gitai2025} tracks which lines of code were written by AI agents, linking every AI-authored line to the agent, model, and original prompt transcripts. Their \texttt{/ask} skill allows agents to query the original intent and requirements behind code they are modifying, providing what the authors describe as access to ``the why'' rather than just ``what the code does.''

Git-ai preserves the \emph{raw transcript}---the full conversation between human and agent. Lore preserves the \emph{distilled knowledge}---the constraints, rejected alternatives, and directives extracted from that conversation. These are related but distinct: the transcript is the raw material, Lore is the refined product. A transcript for a single feature might be thousands of tokens of back-and-forth; the corresponding Lore commit is a dozen structured trailers. For consumption by future agents operating under context window constraints, the compression from transcript to structured knowledge is critical.

\subsection{AI-Generated Commit Messages}

A large ecosystem of tools---including aicommits~\cite{aicommits}, Windsurf's built-in feature~\cite{windsurf}, and Cursor's commit generation~\cite{cursor}---use AI to automatically generate commit messages from diffs. These tools improve consistency and save developer time, but they fundamentally operate as better \emph{labelers} of diffs: they describe what changed, not why. As one practitioner noted, while AI handles ``consistency and technical summaries'' well, it cannot capture ``the `why' behind our changes---the business context, user feedback, or strategic decisions''~\cite{lawson2024}. Lore proposes a categorically different kind of information in commits---not a better description of the diff, but knowledge that was never in the diff at all.

\section{The Lore Protocol}

\subsection{Core Thesis}

A commit should not be a label on a diff. It should be the atomic unit of institutional knowledge in a software project---the smallest self-contained record of a decision, permanently and immutably bound to the code change that enacted it. I call this unit a \textbf{Lore atom}. Its defining properties:

\textbf{Atomic binding.} The knowledge is permanently fused to the exact code change it describes. Unlike documentation, wikis, ADRs, or chat threads, it cannot drift out of sync---the commit and its message are a single immutable object.

\textbf{Temporal immutability.} Once committed, the Lore atom exists permanently in the project history. It forms an append-only log of decisions---a stronger guarantee than any documentation system provides.

\textbf{Universal availability.} Every git-based project already has this channel. There is zero infrastructure cost. The distribution mechanism (clone, fetch, pull) is already solved.

\textbf{Natural granularity.} Commits are already scoped to logical units of work. The Lore atom inherits this scoping for free.

\subsection{Format: Git Trailers}

Lore uses git trailers---a native git feature for structured key-value data in commit messages. This choice is deliberate: no custom parsing is required, the format is queryable through standard git commands (\texttt{git log --trailer=}), and it is supported by every git tool in existence.

A Lore-enriched commit takes the following form (Figure~\ref{fig:commit}):

\begin{figure}[ht]
\begin{lstlisting}
Prevent silent session drops during long-running operations

The auth service returns inconsistent status codes on token
expiry, so the interceptor catches all 4xx responses and
triggers an inline refresh.

Constraint: Auth service does not support token introspection
Constraint: Must not add latency to non-expired-token paths
Rejected: Extend token TTL to 24h | security policy violation
Rejected: Background refresh on timer | race condition
Confidence: high
Scope-risk: narrow
Reversibility: clean
Directive: Error handling is intentionally broad (all 4xx)
  -- do not narrow without verifying upstream behavior
Tested: Single expired token refresh (unit)
Not-tested: Auth service cold-start > 500ms behavior
\end{lstlisting}
\caption{Example of a Lore-enriched commit message.}
\label{fig:commit}
\end{figure}

The \textbf{intent line} (first line) describes \emph{why} the change was made, not what changed. The \textbf{body} provides narrative context. The \textbf{trailers} carry machine-parseable decision context. The trailer vocabulary is shown in Table~\ref{tab:trailers}.

\begin{table}[ht]
\centering
\small
\caption{Lore trailer vocabulary.}
\label{tab:trailers}
\begin{tabular}{@{}lp{7.5cm}@{}}
\toprule
\textbf{Trailer} & \textbf{Semantics} \\
\midrule
\texttt{Constraint:} & Rules that shaped this decision and may still be active \\
\texttt{Rejected:} & Alternatives evaluated and dismissed, with reasons \\
\texttt{Confidence:} & Author's assessment: \texttt{low}, \texttt{medium}, \texttt{high} \\
\texttt{Scope-risk:} & Blast radius: \texttt{narrow}, \texttt{moderate}, \texttt{wide} \\
\texttt{Reversibility:} & \texttt{clean}, \texttt{migration-needed}, \texttt{irreversible} \\
\texttt{Directive:} & Forward-looking instructions for future modifiers \\
\texttt{Tested:} & What was verified and how \\
\texttt{Not-tested:} & What was not verified and why \\
\texttt{Related:} & Reference to related commits by hash \\
\bottomrule
\end{tabular}
\end{table}

Every trailer is optional. The format is additive and extensible: unknown keys are ignored, and teams can introduce domain-specific trailers without breaking compatibility.

\subsection{The CLI: Querying and Authoring Lore}

While raw Lore trailers are readable in any \texttt{git log} output, a CLI tool provides efficient, scoped querying (Figure~\ref{fig:cli}):

\begin{figure}[ht]
\begin{lstlisting}
$ lore --help

Lore -- Query and author institutional knowledge from git history

Query commands:
  lore context <path>       Full lore summary for a code region
  lore constraints <path>   Active constraints shaping this code
  lore rejected <path>      Previously rejected alternatives
  lore directives <path>    Forward-looking warnings
  lore coverage <path>      Test coverage map
  lore stale [--older-than] Flag outdated assumptions

Authoring commands:
  lore commit               Interactive commit builder
  lore commit --from-json   Commit from structured input
  lore validate             Check recent commits for lore format
\end{lstlisting}
\caption{Lore CLI interface.}
\label{fig:cli}
\end{figure}

\subsection{Agent Consumption Model}

When an AI coding agent enters a Lore-enabled project, the consumption workflow unfolds through discoverable CLI interactions:

\textbf{Discovery.} The agent encounters a \texttt{.lore} config file or Lore-formatted commits in \texttt{git log}, and runs \texttt{lore --help} to understand the available commands.

\textbf{Constraint harvest.} Before modifying any file, the agent runs \texttt{lore context <path>} to load the full decision history into its working context.

\textbf{Anti-pattern filtering.} \texttt{Rejected:} trailers prevent the agent from re-exploring dead ends.

\textbf{Directive absorption.} \texttt{Directive:} trailers function as messages from previous authors to future modifiers---permanent, searchable institutional knowledge.

\textbf{Temporal reasoning.} The \texttt{lore stale} command surfaces constraints that may be outdated, enabling a self-healing knowledge base.

\textbf{Knowledge serialization.} The agent formats its own commit as a Lore atom, serializing decision context it already holds rather than discarding it.

The critical property is that \textbf{zero special agent support is required}---only the ability to run shell commands and read text output.

\section{Analysis: Why Not Something Else?}

\textbf{vs.\ ADRs.} Complementary, not competing. ADRs capture architectural decisions; Lore captures implementation decisions. ADRs are separate files subject to synchronization decay; Lore atoms are immutably bound to code changes.

\textbf{vs.\ Code Comments.} Comments are mutable, routinely inaccurate, and cannot capture trajectory. Lore preserves the full ordered history of decisions.

\textbf{vs.\ AI-Generated Diff Summaries.} Diff summaries~\cite{aicommits, windsurf} re-encode information already in the diff. Lore adds information that was \emph{never in the diff}: rejected alternatives, external constraints, confidence assessments, and forward-looking warnings.

\textbf{vs.\ RAG Over External Sources.} Chat-based RAG suffers from low signal-to-noise, no atomic binding, and ephemeral access. Lore inverts all three.

\textbf{vs.\ Code Digital Twin.} The Code Digital Twin~\cite{peng2025cdt} prescribes heavyweight infrastructure. Lore offers near-zero infrastructure cost for the vast majority of projects that cannot justify a full knowledge layer.

\section{Critical Examination}

\textbf{Overhead objection.} Lore is situated in the AI agent era. Agents already hold decision context and can serialize it at commit time; the overhead is review, not authoring. \emph{Residual risk:} rubber-stamped low-quality atoms require \texttt{lore validate} checks.

\textbf{Context window objection.} No context window expansion recovers information never written down. Rejected alternatives and external constraints exist only in the author's mind at commit time. \emph{Concession:} Lore's value is primarily inter-session and inter-agent.

\textbf{Trust and gaming.} Structured format makes quality measurable and auditable. \emph{Concession:} Lore does not solve the trust problem; it makes trust auditable.

\textbf{Granularity mismatch.} The \texttt{Related:} trailer supports commit chains. \emph{Concession:} ADRs~\cite{nygard2011} remain better for cross-cutting architectural decisions.

\section{Architecture}

Lore has exactly two layers. \textbf{Layer A} is the commit format---git trailers in standard messages, requiring zero tooling. \textbf{Layer B} is the CLI tool---providing efficient querying and authoring. The CLI is optional; the format is sufficient. No servers, no index files, no databases. Additional tools can be built on top without becoming part of the core protocol.

\section{Adoption Path and Validation}

Lore degrades gracefully: best experience with the CLI, fallback via native \texttt{git log --trailer=}, and no worse than today in conventional repos. The adoption curve is agent-first: AI agents produce Lore commits with zero friction. Conventional Commits~\cite{conventionalcommits} took years because humans had to type them; Lore shifts the burden to agent configuration.

\emph{Proposed empirical validation:} Two teams, same project, six months. One uses Lore; one uses conventional commits. Measured outcomes: agent task success rate, time-to-correct-solution, rate of re-proposing rejected approaches, and review cycles before merge.

\section{Conclusion}

This paper has argued that the git commit message is critically underutilized as a knowledge channel. I have proposed Lore, a protocol that repurposes commit messages into structured decision records using native git trailers, queryable through a CLI that any AI agent can discover and use.

The core insight is that every commit already produces institutional knowledge that is currently discarded. Lore asks agents and developers to serialize this knowledge rather than throw it away. The overhead is minimal; the infrastructure requirement is zero; and the value compounds over time through a knowledge flywheel that conventional commits cannot produce.

To the best of my knowledge, no prior work has proposed the commit message itself as the vehicle for structured, machine-parseable decision context optimized for AI agent consumption. Lore operates at the simplest possible layer. This minimalism is a feature, not a limitation.

Every codebase has lore. Most of it is lost. The protocol proposed here is a first step toward changing that.

\section*{Disclosure of AI-Assisted Tools}

The author used AI-assisted tools during the research and writing process for this paper. Specifically, Claude (Anthropic, claude-opus-4-6) was used as an interactive research collaborator for brainstorming the core thesis, iterating on protocol design decisions, conducting structured literature review, and drafting and refining text. Google NotebookLM was used for organizing and synthesizing research materials. All intellectual contributions---including the identification of the Decision Shadow problem, the Lore protocol design, the choice of git trailers as the implementation mechanism, the CLI-first architecture, and the comparative analysis against competing approaches---originated from the author's own reasoning, with AI tools serving as instruments for exploration, articulation, and refinement. The author reviewed, edited, and takes full responsibility for all content in this paper.


\begin{thebibliography}{16}

\bibitem{peng2025cdt}
X.~Peng and C.~Wang, ``Code Digital Twin: Empowering LLMs with Tacit Knowledge for Complex Software Development,'' \emph{arXiv preprint arXiv:2503.07967}, 2025.

\bibitem{terragni2025future}
V.~Terragni, A.~Vella, P.~Roop, and K.~Blincoe, ``The Future of AI-Driven Software Engineering,'' \emph{ACM Trans.\ Softw.\ Eng.\ Methodol.}, 2025.

\bibitem{gcc2025}
``Git Context Controller: Manage the Context of LLM-based Agents like Git,'' \emph{arXiv preprint arXiv:2508.00031}, 2025.

\bibitem{nygard2011}
M.~Nygard, ``Documenting Architecture Decisions,'' November 2011. \url{https://cognitect.com/blog/2011/11/15/documenting-architecture-decisions}

\bibitem{conventionalcommits}
``Conventional Commits Specification v1.0.0.'' \url{https://www.conventionalcommits.org/}

\bibitem{claudecode}
Anthropic, ``Claude Code,'' 2025. \url{https://docs.anthropic.com/en/docs/agents-and-tools/claude-code/overview}

\bibitem{copilot}
GitHub, ``GitHub Copilot,'' 2025. \url{https://github.com/features/copilot}

\bibitem{cursor}
Anysphere, ``Cursor: The AI Code Editor,'' 2025. \url{https://cursor.com}

\bibitem{gitclear2024}
B.~Harding, ``Coding on Copilot: 2023 Data Suggests Downward Pressure on Code Quality,'' GitClear, 2024.

\bibitem{18f2021}
18F (U.S.\ General Services Administration), ``Architecture Decision Records: Helpful now, invaluable later,'' July 2021.

\bibitem{haylock2025}
D.~Haylock, ``The Importance of Architecture Decision Records (ADRs),'' \emph{Medium}, March 2025.

\bibitem{log4brains}
T.~Vaill, ``Log4brains: ADR management and publication tool.'' \url{https://github.com/thomvaill/log4brains}

\bibitem{gitai2025}
A.~Cunniffe, ``Git AI: A Git extension for tracking AI-generated code,'' 2025. \url{https://usegitai.com}

\bibitem{aicommits}
H.~Nasser, ``aicommits: A CLI that writes your git commit messages with AI.'' \url{https://github.com/Nutlope/aicommits}

\bibitem{windsurf}
Windsurf, ``AI Commit Messages,'' 2025. \url{https://docs.windsurf.com/windsurf/ai-commit-message}

\bibitem{lawson2024}
C.~Lawson, ``Git Commit: When AI Met Human Insight,'' \emph{Versent Tech Blog}, September 2024.

\end{thebibliography}
\end{document}